\documentclass[a4paper,11pt]{article}
\usepackage{pos}
\newcommand{\Dp}{{\bf \Delta}_{\perp}}
\newcommand{\be}{\begin{eqnarray}}
	\newcommand{\ee}{\end{eqnarray}}

\newcommand{\bfR}{{\bf R}_{\perp}}
\newcommand{\bfz}{{\bf 0}_{\perp}}

\newcommand{\bfk}{{\bf k}_{\perp}}

\newcommand{\bfki}{{\bf k}_{\perp i}}

\newcommand{\bfb}{{\bf b}_{\perp}}

\newcommand{\bfP}{{\bf P}_{\perp}}

\newcommand{\bfp}{{\bf p}_{\perp}}

\newcommand{\bfr}{{\bf r}_{\perp}}
\title{Single spin asymmetry among hyperons using scalar diquark model}

\author[a]{Navpreet Kaur}
\author*[a]{Harleen Dahiya}

\affiliation[a]{Department of Physics, Dr. B.R. Ambedkar National
	Institute of Technology, Jalandhar, 144008, India}

\emailAdd{knavpreet.hep@gmail.com}
\emailAdd{dahiyah@nitj.ac.in}

\abstract{The Fourier transformed chiral even generalized parton distributions $(GPDs)$ for purely transverse momentum transfer $(\Delta^+=0)$ characterize an individual parton distribution of a hyperon in a perpendicular plane at some distance from the center of momentum of the system. Further, the presence of left-right asymmetry in the distorted parton distribution gives a clue for the presence of single-spin asymmetry. In order to have a deep insight into the distortions, we have exhibited the information about asymmetries from the spin flip matrix element of a $GPD$ for hyperons by employing scalar diquark model and also compared parton distributions of different possible combinations of quark-diquark pair to get substantial structural information on hyperons. }

\FullConference{The XVIth Quark Confinement and the Hadron Spectrum Conference (QCHSC24)\\
 19-24 August, 2024\\
 Cairns Convention Centre, Cairns, Queensland, Australia\\}

\begin{document}
\maketitle

\section{Introduction}
Partonic information with a longitudinal momentum fraction $x$ at a specific transverse position inside the hadron can be studied through the generalized parton distributions (GPDs). Deeply virtual Compton scattering and deeply virtual meson production are the processes through which these distributions can be analyzed experimentally and computed theoretically as a non-forward matrix element of the light-cone correlator \cite{Radyushkin:1997ki}. GPDs are defined in terms of longitudinal momentum fraction $x$, skewness parameter $\zeta$ and the invariant momentum transfer $-t=\Dp^2$. For zero skewness, the GPDs correspond to purely transverse momentum transfer. In impact parameter space, Fourier transformed GPDs interpret the spatial distribution of partons in the plane transverse to the momentum of a fast moving hadron \cite{Burkardt:2002ks}. The center of momentum of the hadron target in transverse plane is treated as the reference point for the impact parameter space distributions. The physical significance of the
spin-flip GPD $E_{X}^{q}(x,0,t)$ in the infinite momentum frame can be understood in terms of the distorted partonic distributions of $q$ flavored quarks in the transverse plane when the target state is transversely polarized. The existence of left-right asymmetries in the parton distributions can be associated with the single-spin asymmetry (SSA) by introducing the concept of final state interaction (FSI) \cite{Burkardt:2003je}. SSA is time-reversal odd, appears in
quantum chromodynamics (QCD) through phase difference in the different amplitudes of spin, and is measured in semi-inclusive deep inelastic scattering. A QCD-inspired scalar diquark model of a hadron has been used to analyze the transverse distortion and SSA of $\Sigma^+$ and $\Xi^o$ hyperons as it has all the Lorentz symmetries and gives a simple view of the quark-diquark system.
\section{Scalar diquark model}
We have considered working with the light-cone dynamics in which the fields are quantized at a fixed light-cone time $\tau=t_o +z/c$, instead of an ordinary time $t_o$. A hadron state with momentum $P$ can be expanded in terms of its $\mathcal{N}$ constituent partons as 
\begin{eqnarray}
	|\mathcal{B}(P^{+},{\bf P_\perp^2)}\rangle &=& \sum_{\mathcal{N}} \prod_{i=1}^{\mathcal{N}} \frac{dx_i~  d^2\bfki}{2(2\pi)^3\sqrt{x_{i}}} \, 16 \pi^{3} \, \delta \bigg(1-\sum_{i=1}^{\mathcal{N}} x_{i}\bigg) \, \delta^{(2)} \bigg(\sum_{i=1}^{\mathcal{N}}\bfki\bigg) \nonumber \\		
	&\times& \psi_{\mathcal{N}}(x_{i},\bfki,\lambda_{i})|\mathcal{N}; x_{i} P^{+},x_{i}\bfP+\bfki,\lambda_{i}\rangle \, ,
\end{eqnarray}
where $x_i$, $\bfki$ and $\lambda_i$ correspond to the light-cone longitudinal momentum fraction, intrinsic transverse momentum and helicity carried by \textit{i}th constituent of a baryon, respectively. The momentum carried by each constituent parton with mass $m_i$ of a hadron in the form of light-cone coordinates can be expressed as
\begin{equation}
	k_{i}=(k_{i}^{+},k_{i}^{-},\bfki)=\bigg(x_{i}P^{+},\frac{\bfki^{2}+m_{i}^{2}}{x_{i}P^{+}},\bfki\bigg) \, .
\end{equation}
There exists only two possible spin combinations for the two-particle Fock state, depending upon one-loop quantum fluctuations of the Yukawa theory. For a hadron with total spin $Jz=+\frac{1}{2} (\Uparrow)$, it can be written as
\be
|\psi_{2 particle}^{\Uparrow X}(P^{+},\bfP=\bf{0}_{\perp})\rangle & =& \int \frac{dx~  d^2\bfki}{2(2\pi)^3\sqrt{x(1-x)}} \, \bigg[\psi_{+\frac{1}{2}}^{\Uparrow X}(x,\bfk)\bigg|+\frac{1}{2};x P^{+},\bfk\bigg\rangle \nonumber \\
&+& \bigg[\psi_{-\frac{1}{2}}^{\Uparrow X}(x,\bfk)\bigg|-\frac{1}{2};x P^{+},\bfk\bigg\rangle \bigg] \, .
\ee
With the application of Fourier transformation \cite{Kim:2008ghb}, momentum space can be swapped to the impact parameter space as follows
\be
\psi(x,\bfb) = \frac{1}{1-x} \int \frac{d^2 \bfk}{4\pi^{2}} e^{\frac{i \, \bfk \cdot \bfb}{1-x}} \psi (x,\bfk) \, .
\ee
Here, $\bfb=b_{\perp}(\cos \phi_{b}, \,\sin \phi_{b})$. Hence, the light-cone wave functions (LCWFs) in terms of its impact parameter space coordinates can be described as \cite{Lorce:2017wkb}
\be
\psi_{+\frac{1}{2}}^{\Uparrow X}(x,\bfb) &=& -\frac{g}{2 \pi} \frac{x M_{X}+m_{q}}{\sqrt{1-x}} K_{0}(Z) \, ,  \\
\psi_{-\frac{1}{2}}^{\Uparrow X}(x,\bfb) &=& \frac{ig}{2 \pi} \frac {\sqrt{\mathcal{M}_{X}^{qn}(x)} e^{i \phi_{b}}}{\sqrt{1-x}} K_{1}(Z) \, ,
\ee
where $M_X$,  $m_q$ and $\mu_n$ represent the masses of $X$ baryon, $q$ ($u/s$) flavored quark and diquark $n$ ($uu/us/ss$), respectively. The terms $Z$ and $\mathcal{M}_{X}^{qn} (x) $ have the mathematical form, $Z=\sqrt{\mathcal{M}_{X}^{qn} (x)}|\bfb|/(1-x)$ and $\mathcal{M}_{X}^{qn} (x) = m_{q}^{2}(1-x)+\mu^{2}_{n}x-M^{2}_{X}x(1-x)$ with $K_{p}$ as the modified Bessel function of the second kind $p$. In a similar way, for a hadron with total spin $Jz=-\frac{1}{2} (\Downarrow)$, it can be written as
\be
|\psi_{2 particle}^{\Downarrow X}(P^{+},\bfP=\bf{0}_{\perp})\rangle & =& \int \frac{dx~  d^2\bfki}{2(2\pi)^3\sqrt{x(1-x)}} \, \bigg[\psi_{+\frac{1}{2}}^{\Downarrow X}(x,\bfk)\bigg|+\frac{1}{2};x P^{+},\bfk\bigg\rangle \nonumber \\
&+& \bigg[\psi_{-\frac{1}{2}}^{\Uparrow X}(x,\bfk)\bigg|-\frac{1}{2};x P^{+},\bfk\bigg\rangle \bigg] \, ,
\ee
and in terms of the impact parameter space coordinates, these momentum space LCWFs can be described as $\psi_{+\frac{1}{2}}^{\Downarrow X}(x,\bfb) = \psi_{+\frac{1}{2}}^{\Uparrow X}(x,\bfb)$ and $\psi_{-\frac{1}{2}}^{\Downarrow X}(x,\bfb) = [\psi_{-\frac{1}{2}}^{\Uparrow X}(x,\bfb)]^\ast$.


\section{Transverse distortion of the parton distributions}
\begin{figure*}
\centering
\begin{minipage}[c]{0.98\textwidth}
(a)\includegraphics[width=5.7cm]{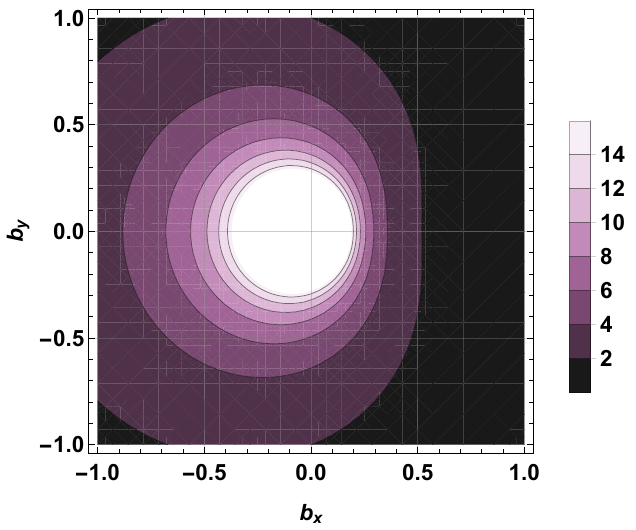}
\hspace{0.03cm}
(b)\includegraphics[width=5.7cm]{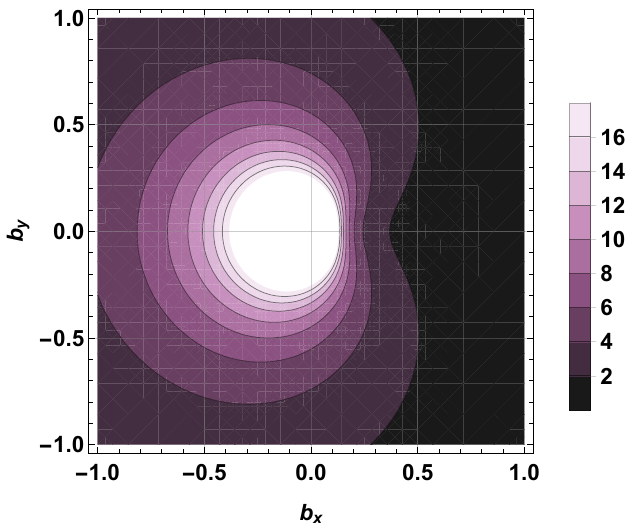}
\hspace{0.03cm} \\
(c)\includegraphics[width=5.7cm]{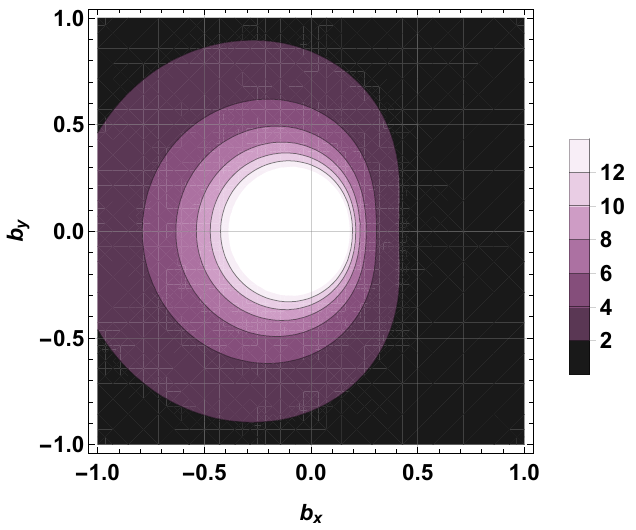}
\hspace{0.03cm}
(d)\includegraphics[width=5.7cm]{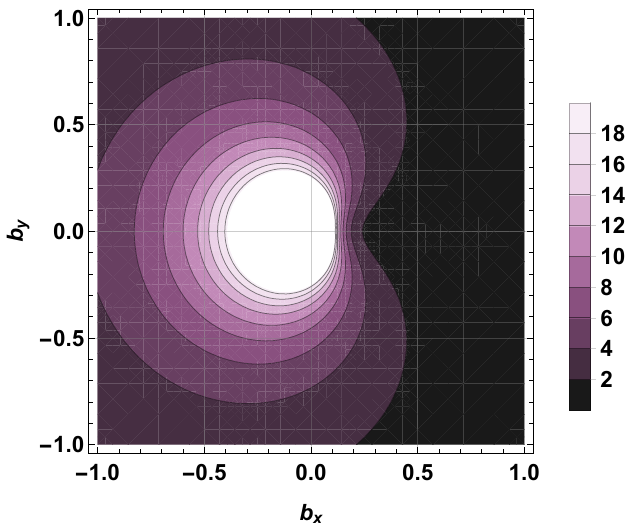}
\hspace{0.03cm}
\end{minipage}
\caption{\label{FigSigma} (Color online) Transverse distortion for $u$ (row 1) and $s$ (row 2) quark flavor in $\Sigma^{+}$ for longitudinal momentum fraction $ x=0.1$ (column 1) and $0.3$ (column 2) respectively. } 
\end{figure*}

GPDs give a substantial spatial information of the partons inside a hadron. The matrix elements of the bilinear vector currents provide quark helicity independent, chiral even GPDs and the correlator to obtained these GPDs can be written as \cite{Brodsky:2000xy}
\be
\frac{1}{2}\int \frac{dy^{-}}{2 \pi} &e^{ix \bar{P}^{+}y^{-}}& \bigg\langle P^{\prime}\bigg|\bar{\psi}\bigg(\frac{-y}{2}\bigg) \,\gamma^{+} \,\psi\bigg(\frac{y}{2}\bigg)\bigg|P\bigg\rangle \bigg|_{y^{+}=0, \bf{y_{\perp}}=0} \nonumber \\ &=& \frac{1}{2 \bar{P^+}}\bar{u}(P^\prime)\bigg[H_{X}^{q}(x,0,t) \,\gamma^{+}+E_{X}^{q}(x,0,t) \,\frac{i \sigma^{+\alpha}(-\Delta_{\alpha})}{2M_{X}}\bigg] u(P) \, ,
\label{GPDcorr}
\ee
where $u(P)$ and $\bar{u}(P^\prime)$ are the light-cone spinors of the initial and final stae of baryons. The average momentum of the initial and final state of a baryon is denoted by $\bar{P}$. 
A distortion in the parton distributions can be investigated for a transversely polarized baryon target, if a spin-flip GPD for the considered target exists. We have considered $E^{q}_{X}(x,\zeta=0,t)$ GPD that involves the helicity flip amplitude of a baryon to interpret the distortion of the parton distributions. Hence, to investigate its density interpretation, consider a state, polarized in the $y$-direction in the infinite momentum frame
\be
|Y\rangle = \frac{1}{\sqrt{2}} \big[|p^{+},\bfR=\bfz,\uparrow\rangle + i \, |p^{+},\bfR=\bfz,\downarrow\rangle \big] \, ,
\ee
where
\be
|p^{+},\bfR=\bfz,\lambda\rangle = N \int d^{2} \bfp |p^{+},\bfp,\lambda\rangle \, .
\ee
The unpolarized quark distribution for the given defined state can be expressed in terms of the impact parameter space coordinates as 
\be
q_{\hat{y}}^{X_{q}} (x,\bfb) &=& \int \frac{d^{2} \Dp }{(2 \pi)^{2}} e^{-i \Dp \cdot \bfb} \bigg[ H_{X}^{q}(x,0,t) + i \frac{\Dp^{x}}{2M_{X}} E_{X}^{q}(x,0,t) \bigg] \nonumber \\
&=&\mathcal{H}_{X}^{q} (x,\bfb)+ \frac{1}{2M_{X}} \frac{\partial}{\partial b^{x}} \mathcal{E}_{X}^{q} (x,\bfb) \, .
\label{transdistort}
\ee
In the present calculations, we have used masses of an active quark, spectator diquark and hyperons as input parameters in the units of GeV. Their numerical values are $m_u=0.33$, $m_s=0.48$, $\mu_{uu}=0.8$, $\mu_{us}=0.95$, $\mu_{ss}=1.10$, $M_{\Sigma^+}=1.189$ and $M_{\Xi^o}=1.192$. The term expressed in Eq. (\ref{transdistort}) has been plotted in Fig. \ref{FigSigma} for the $u$ quark flavor of $\Sigma^+$ in the row $1$ for (a) $x=0.1$ and (b) $x=0.3$. We found that as the longitudinal momentum fraction increases, the distribution get more distorted. This similar pattern is followed by the $s$ quark flavor $\Sigma^+$, presented in Fig. \ref{FigSigma} for (c) $x=0.1$ and (d) $x=0.3$. However, $s$ quark flavor is concentrated in smaller region for $x=0.1$ than $u$ quark which attributes to the less tendency of carrying smaller longitudinal momentum fraction by the $s$ quark flavor. At $x=0.3$, comparison between the transverse distortion plots presented in Fig. \ref{FigSigma} (column $2$) show that $s$ quark flavor experiences more transverse distortion than $u$ quark flavor. \par
\begin{figure*}
\centering
\begin{minipage}[c]{0.98\textwidth}
(a)\includegraphics[width=5.7cm]{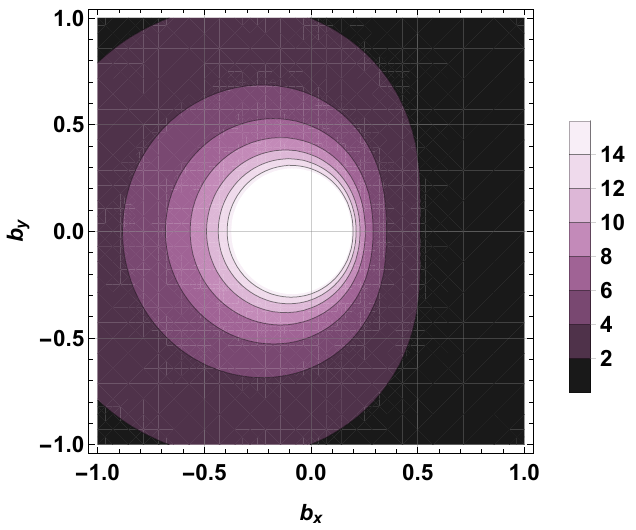}
\hspace{0.03cm}
(b)\includegraphics[width=5.7cm]{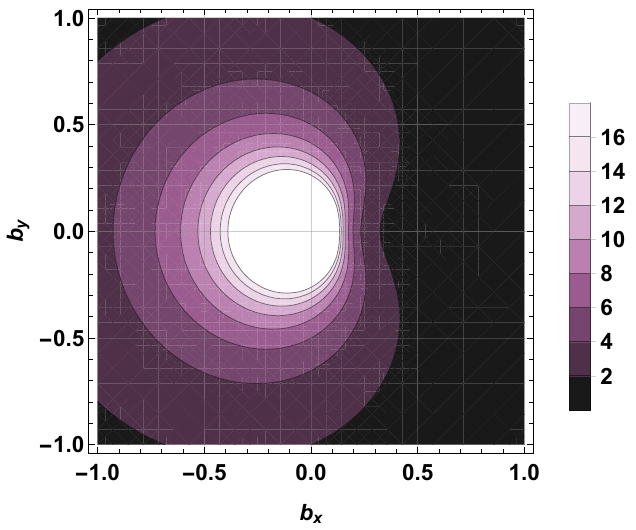}
\hspace{0.03cm} \\
(c)\includegraphics[width=5.7cm]{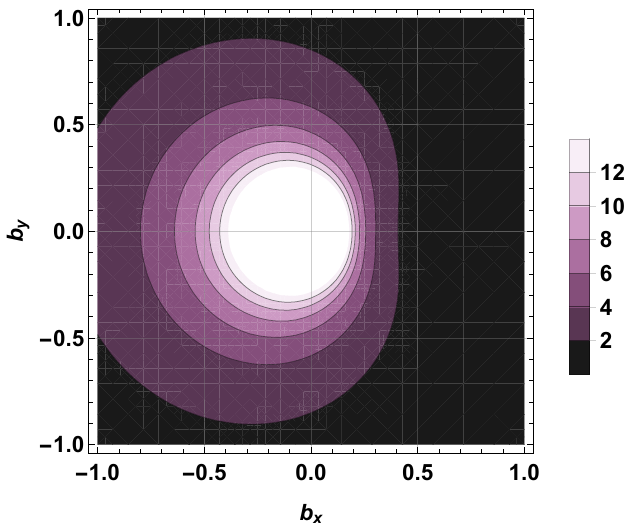}
\hspace{0.03cm}
(d)\includegraphics[width=5.7cm]{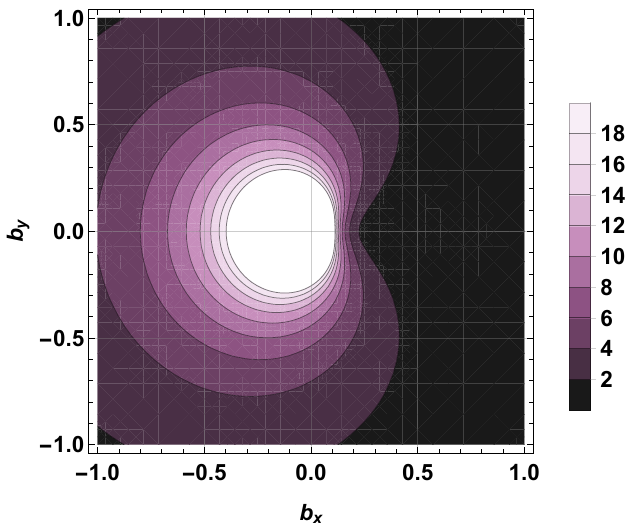}
\hspace{0.03cm}
\end{minipage}
\caption{\label{FigXi} (Color online) Transverse distortion for $u$ (row 1) and $s$ (row 2) quark flavor in $\Xi^{o}$ for longitudinal momentum fraction $ x=0.1$ (column 1) and $0.3$ (column 2) respectively. } 
\end{figure*}

Fig. \ref{FigXi} represent the plots of terms expressed in Eq. (\ref{transdistort}) for the case of $u$ quark flavor of $\Xi^o$ for (a) $x=0.1$ and (b) $x=0.3$. These plots indicate that with an increase in longitudinal momentum fraction from $0.1$ to $0.3$, more pronounced transverse distortion is observed but with smaller spread in the transverse plane as compared to spread of transverse distortion obtained for the case of $\Sigma^+$. This smaller spread is due to the dominance of heavier spectator diquark $ss$ in $\Xi^o$. At $x=0.1$, comparison between the transverse distortion plots presented in Fig. \ref{FigXi} (column $1$) of $\Xi^o$ ensure that lighter quark flavor has capacity of carrying smaller longitudinal momentum fraction as the spread of distribution is more for $u$ quark flavor. Comparison between the transverse distortion plots presented in Fig. \ref{FigXi} (column $2$) of $\Xi^o$ at $x=0.3$ shows more transverse distortion and spread for $s$ quark flavor of $\Xi^o$ because the diquark $us$ becomes lighter. The presence of left-right asymmetry gives a hint for the presence of SSA as they are connected by the relation SSA $=$ GPD $\ast$ FSI. 
\section{Single-spin asymmetry}
\begin{figure*}
\centering
\begin{minipage}[c]{0.98\textwidth}
(a)\includegraphics[width=7cm]{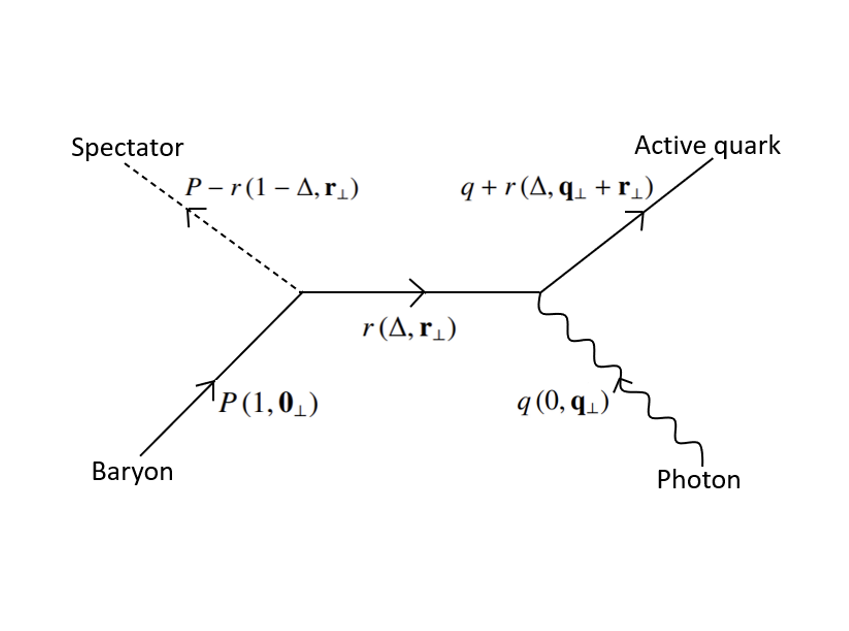}
\hspace{0.03cm}
(b)\includegraphics[width=7cm]{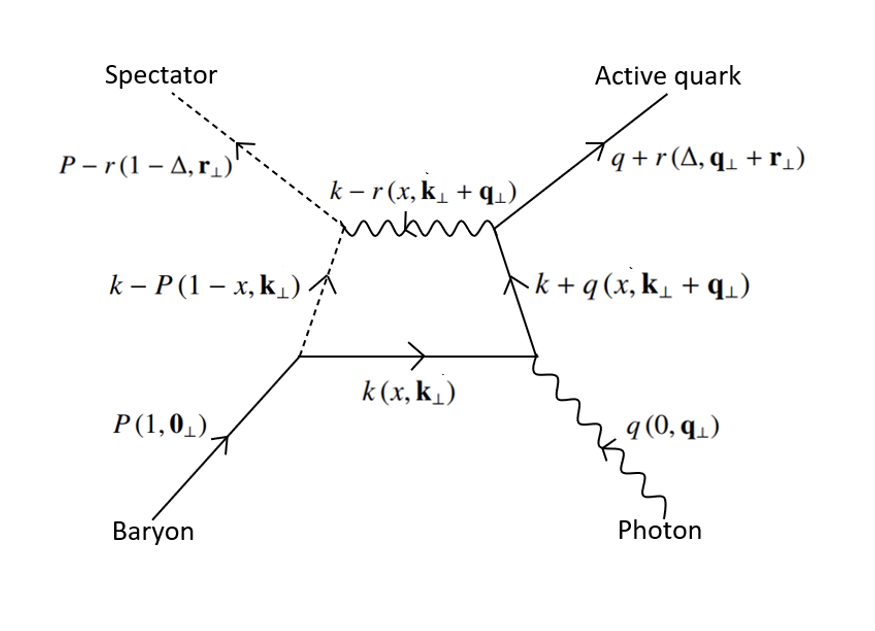}
\hspace{0.03cm} 
\end{minipage}
\caption{\label{FigFeyn} (Color online) Tree and one-loop Feynman diagrams for $X+\gamma^{\ast} \rightarrow q + \mu_{n}^0$. } 
\end{figure*}

FSI occurring in the semi-inclusive deep inelastic scattering, takes place between  an active quark that has already interacted with a virtual photon and its unaffected spectator. In order to produce SSA, the indispensable phase can be incorporated by the insertion of FSI and interference among the amplitudes of a process $X+\gamma^{\ast} \rightarrow q + \mu_{n}^0$, the superscript on the diquark mass denote its zero spin. Fig. \ref{FigFeyn} represents the tree and one-loop Feynman graphs that can be employed to obtain amplitude of the mentioned process.
The SSA can be expressed in terms of the amplitudes as \cite{Brodsky:2002cx}
\be
\mathcal{P}_{y}^{X(q)}=&\frac{1}{\mathcal{C}}& \big(i \, (\mathcal{A}_{X}^{q}(\Uparrow \rightarrow \uparrow)^{\ast} \, \mathcal{A}_{X}^{q}(\Downarrow \rightarrow \uparrow)-\mathcal{A}_{X}^{q}(\Uparrow \rightarrow \uparrow) \, \mathcal{A}_{X}^{q}(\Downarrow \rightarrow \uparrow)^{\ast}) \nonumber \\
&+& i \, (\mathcal{A}_{X}^{q}(\Uparrow \rightarrow \downarrow)^{\ast} \, \mathcal{A}_{X}^{q}(\Downarrow \rightarrow \downarrow)-\mathcal{A}_{X}^{q}(\Uparrow \rightarrow \downarrow) \, \mathcal{A}_{X}^{q}(\Downarrow \rightarrow \downarrow)^{\ast}) \big) \, ,
\ee
where the normalization $\mathcal{C}$ for the unpolarized cross section is given by
\be
\mathcal{C}=|\mathcal{A}_{X}^{q}(\Uparrow \rightarrow \uparrow)|^{2} + |\mathcal{A}_{X}^{q}(\Downarrow \rightarrow \uparrow)|^{2} +|\mathcal{A}_{X}^{q}(\Uparrow \rightarrow \downarrow)|^{2} +|\mathcal{A}_{X}^{q}(\Downarrow \rightarrow \downarrow)|^{2} .
\ee
Fig. \ref{FigFeyn} presents the tree and one-loop Feynman graphs that have been employed to get amplitude of the above mentioned process. The structure of these amplitudes is
\be
\mathcal{A}_{X}^{q}(\Uparrow \rightarrow \uparrow)&=&\bigg(M_{X}+\frac{m_{q}}{\Delta}\bigg) \, C \bigg(h_{X} +i \frac{e_{1}e_{2}}{8 \pi} I_{1X}\bigg) \, , \label{Amp1} \\
\mathcal{A}_{X}^{q}(\Downarrow \rightarrow \uparrow)&=&\bigg(\frac{r_{1}-i \, r_{2}}{\Delta}\bigg) \, C \bigg(h_{X} +i \frac{e_{1}e_{2}}{8 \pi} I_{2X}\bigg) \,, \label{Amp2} \\
\mathcal{A}_{X}^{q}(\Uparrow \rightarrow \downarrow)&=&\bigg(-\frac{r_{1}+i \, r_{2}}{\Delta}\bigg) \, C \bigg(h_{X} +i \frac{e_{1}e_{2}}{8 \pi} I_{2X}\bigg) \,, \label{Amp3} \\
\mathcal{A}_{X}^{q}(\Downarrow \rightarrow \downarrow) &=& \bigg(M_{X} + \frac{m_{q}}{\Delta}\bigg) \, C \bigg(h_{X} +i \frac{e_{1}e_{2}}{8 \pi} I_{1X}\bigg) \, . \label{Amp4}
\ee
We have
\be
C&=&-g \, e_{1} P^{+} \sqrt{\Delta} \, 2 \, \Delta (1-\Delta) \, , \\
h_{X}&=&\frac{1}{\bfr+\mathcal{M}_{X}^{un}(\Delta)} \, ,\\
I_{1X}&=&\int_{0}^{1} d\alpha \frac{1}{\alpha (1-\alpha)\bfr^{2} + \alpha \mu_{g}^{2}+(1-\alpha) \mathcal{M}_{X}^{un}(\Delta)} \, , \\
I_{2X}&=&\int_{0}^{1} d\alpha \frac{\alpha}{\alpha (1-\alpha)\bfr^{2} + \alpha \mu_{g}^{2}+(1-\alpha) \mathcal{M}_{X}^{un}(\Delta)} \, ,
\ee
where $e_{1}$ and $e_{2}$ are the electric charges of an active flavor of quark and a scalar spectator diquark respectively. Coupling constant of the baryon- active quark- scalar spectator vertex is denoted by $g$. In the structure of amplitudes, the first term is an outcome of the Born contribution of tree representation whereas the second term reflects the one-loop representation that contains two different contributions $I_{1X}$ and $I_{2X}$. When $Q^2$ and photon laboratory energy is large with fixed $\Delta=x_{bj}$ in the Bjorken scaling limit, the light-cone laboratory frame and usual laboratory frame becomes same. \par

\begin{figure*}
\centering
\begin{minipage}[c]{0.98\textwidth}
(a)\includegraphics[width=5.7cm]{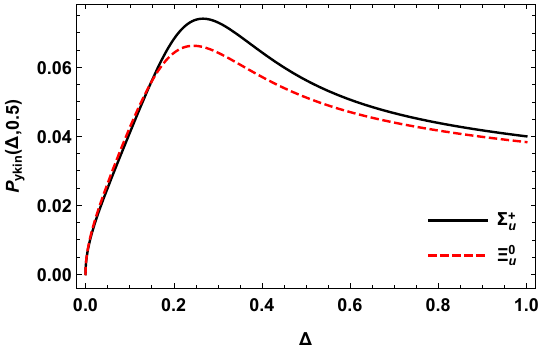}
\hspace{0.03cm}
(b)\includegraphics[width=5.7cm]{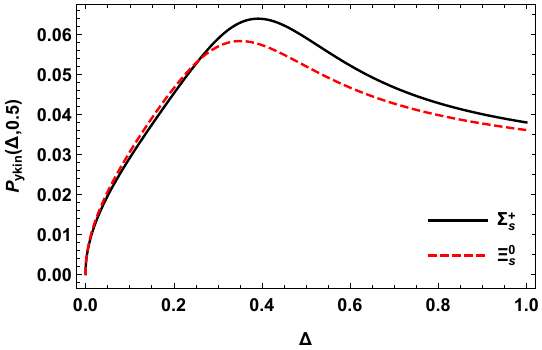}
\hspace{0.03cm} 
\end{minipage}
\caption{\label{FigSSASigma} (Color online) Azimuthal single-spin asymmetry as a function of $\Delta$ for constituent (a) $u$ and (b) $s$ quark flavors of hyperons. } 
\end{figure*}

\begin{figure*}
\centering
\begin{minipage}[c]{0.98\textwidth}
(a)\includegraphics[width=5.7cm]{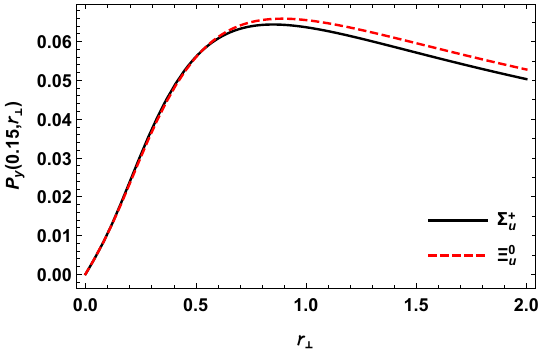}
\hspace{0.03cm}
(b)\includegraphics[width=5.7cm]{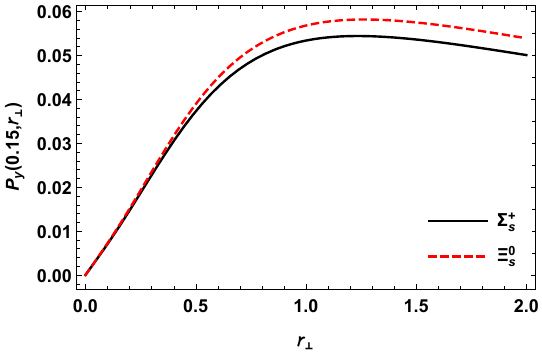}
\hspace{0.03cm} 
\end{minipage}
\caption{\label{FigSSAXi} (Color online) Azimuthal single-spin asymmetry as a function of $|\bfr|$ for constituent (a) $u$ and (b) $s$ quark flavors of hyperons. } 
\end{figure*}

The transverse azimuthal spin asymmetry $\mathcal{P}_y$ can be compared with $A^{sin \, \phi}_{UT}$ which is an experimentally measured HERMES transverse asymmetry observable \cite{Brodsky:2004hh} and the longitudinal asymmetry $\mathcal{P}_y$ can be written as $A^{sin \, \phi}_{UL} \, = \, K \, A^{sin \, \phi}_{UT}$,
where $K$ is a kinematical factor \cite{HERMES:2020ifk}. The outcomes of the longitudinal asymmetry $A^{sin \, \phi}_{UL}$ for hyperons are presented in Figs. \ref{FigSSASigma} and \ref{FigSSAXi} as a function of quark light-front momentum fraction $\Delta$ and the magnitude of active quark momentum jet $\bfr$ relative to the virtual photon direction, respectively, for the constituent quark flavors. The azimuthal single-spin asymmetry as a function of $\Delta$ for the constituent quark flavors of hyperons shows that the lighter $u$ quark flavor has a greater tendency to exhibit SSA as its magnitude is found to be more than the $s$ quark flavor of both hyperons. SSA also attain its peak value at a higher value of $\Delta$ for $s$ quark flavor compared to $u$ quark flavor of the same hyperon. Tendency of exhibiting azimuthal single-spin asymmetry as a function of $\bfr$ for the constituent quark flavors of hyperons show similar trend as shown with respect to $\Delta$. However, the magnitude of SSA as a function of $\bfr$ is found to be more for $\Sigma^+$ than $\Xi^o$. At large values of $\bfr$, the SSA fall off slower for $s$ quark flavors of both hyperons. 
\section{Summary}
We have studied the transverse distortion and single spin asymmetry of $\Sigma^+$ and $\Xi^o$ hyperons to analyze their dependence on different quark flavors. Chiral even GPDs have been evaluated in the impact parameter space to map the distribution of unpolarized quarks. Spin flip GPD $E_X^q(x,0,t)$ measures the transverse deformation in an active quark distribution in the transverse plane. The distortions observed by the polarized state of hyperons show that the lighter $u$ quark flavor has more distribution spread at smaller longitudinal momentum fraction $x=0.1$ than $s$ quark flavor. Whereas at higher longitudinal momentum fraction $x=0.3$, a shrinked distribution has been observed for the lighter $u$ quark flavor. It implies that lighter quark flavors generally carry low longitudinal momentum fraction. The distortion is found to be more for the case of comparatively heavier $s$ quark flavor, irrespective of the longitudinal momentum fraction in both hyperons. The asymmetric distributions observed for the polarized state of hyperons interpret the existence of single-spin asymmetries, measurable in semi-inclusive deep inelastic scattering. In order to relate theoretical predictions with experimental observable, we demonstrate the azimuthal single-spin asymmetry that is accessible in scattering processes in HERMES and JLab experiment projects for different kinds of hadron targets. The distribution of SSA reassures the tendency to carry a higher longitudinaal momentum fraction by a comparatively heavier quark flavor, since the peak of SSA distribution lies on larger values of $\Delta$ and $\bfr$.
\section{Acknowledgment}
H.D. would like to thank  the Science and Engineering Research Board, Anusandhan-National Research Foundation, Government of India under the scheme SERB-POWER Fellowship (Ref No. SPF/2023/000116) for financial support.

\end{document}